\documentclass[letter]{aa} 
\usepackage{graphicx}
\usepackage{txfonts}

\newcommand\E[1]{\times10^{#1}}

\newcommand\SIMEQ{\cong}

\newcommand\U[1]{{\,\rm #1}}

\newcommand\ergs{erg\,s^{-1}}

\newcommand\masyr{mas\,yr^{-1}}
\newcommand\kms{km\,s^{-1}}

\newcommand\al{\alpha}

\newcommand\gm{\gamma}
\newcommand\dl{\delta}
\newcommand\eps{\epsilon}

\newcommand\sg{\sigma}

\newcommand\Om{\Omega}

\newcommand\DlPhi{\Delta\Phi}

\newcommand\rs[1]{_\mathrm{#1}}

\newcommand\dscal{\,(d/1.86\U{kpc})}
\newcommand\dscaldis{\left(\frac{d}{1.86\U{kpc}}\right)}

\newcommand\psr{\ast}
\newcommand\Bpsr{B\rs{\psr}}
\newcommand\Vpsr{V\rs{\psr}}
\newcommand\Vpsrt{V\rs{\psr,t}}

\newcommand\Edot{\dot E}

\newcommand\rhoamb{\rho\rs{amb}}

\newcommand\gmMPD{\gamma\rs{MPD}}

\newcommand\bow{bow}
\newcommand\Rbow{R\rs{\bow}}
\newcommand\Bbow{B\rs{\bow}}

\newcommand\Wfea{W\rs{fea}}
\newcommand\Lfea{L\rs{fea}}
\newcommand\Llif{L\rs{lif}}
\newcommand\EX{E\rs{X}}

\newcommand\Escaldis{\left(\frac{\EX}{2\U{keV}}\right)}

\newcommand\gmX{\gamma\rs{X}}

\newcommand\Bscaldis{\left(\frac{B}{45\U{\mu G}}\right)}
\newcommand\Rgyr{R\rs{gyr}}
\newcommand\Rgyrbow{R\rs{gyr,bow}}
\newcommand\Rgyrfea{R\rs{gyr,fea}}
\newcommand\lmbpar{\lambda_\parallel}

\newcommand\tlif{t\rs{lif}}

\begin{document}
   \title{On the X-ray feature associated with the Guitar Nebula}


   \author{Rino Bandiera}

   \offprints{R. Bandiera}

   \institute{INAF - Osservatorio Astrofisico di Arcetri
              Largo E. Fermi 5, I-50125 Firenze, Italy\\
              \email{bandiera@arcetri.astro.it}
             }

   \date{Received 24 July 2008 / accepted 8 September 2008}

 
  \abstract
{A mysterious X-ray nebula, showing a remarkably linear geometry, was recently
discovered close to the Guitar Nebula, the bow-shock nebula associated with
B2224+65, which is the fastest pulsar known. The nature of this X-ray feature
is unknown, and even its association with pulsar B2224+65 is unclear. }
{We attempt to develop a self-consistent scenario to explain the complex
phenomenology of this object.}
{We assume that the highest energy electrons accelerated at the
termination shock escape from the bow shock and diffuse into the ambient
medium, where they emit synchrotron X-rays. The linear geometry should
reflect the plane-parallel geometry of its ambient field. }
{We estimate the Lorentz factor of the X-ray emitting electrons and the
strength of the magnetic field. The former ($\simeq10^8$) is close to
its maximum possible value, while the latter, at $\SIMEQ45\U{\mu G}$, is
higher than typical interstellar values and must have been amplified in some
way. The magnetic field must also be turbulent to some degree to trap the
electrons sufficiently for synchrotron X-ray emission to occur effectively. We
propose a self-consistent scenario in which, by some streaming instability,
the electrons themselves generate a turbulent field in which they then
diffuse. Some numerical coincidences are explained, and tests are proposed
to verify our scenario.}
{Electron leaking may be common in the majority of pulsar bow-shock nebulae,
even though the X-ray nebulosity in general is too diffuse to be detectable.}

   \keywords{Stars: pulsars: individual: B2224+65 - ISM: individual objects:
Guitar Nebula - Radiation mechanisms: non-thermal}

   \maketitle
%

\section{Introduction}

Pulsar B2224+65 and its bow-shock nebula (the so-called``Guitar Nebula'')
are quite peculiar and extreme objects.
The pulsar is characterized by an exceptionally high velocity.
Its proper motion $\mu=182\U{\masyr}$ translates into a transverse velocity
$\Vpsrt=1,609\dscal\U{\kms}$, where $1.86\U{kpc}$ is the distance estimated
from the pulsar dispersion measure (according to Cordes \& Lazio \cite{cl02}),
which implies that it is the fastest pulsar known.
Otherwise, PSR~B2224+65 behaves like a standard radio pulsar with a
period $P=0.68\U{s}$ and a spin-down power $\Edot=1.20\E{33}\U{\ergs}$,
corresponding to a characteristic time $\tau=1.12\E{6}\U{yr}$ and a pulsar
equatorial field $\Bpsr=2.60\E{12}\U{G}$.
The pulsar timing does not present any anomaly.
Visible in Balmer lines, the nebula displays a peculiar shape (as indicated by
its nickname), which does not resemble in any way a ``well-behaved'' bow shock.
It presents instead a conical head, an elongated neck, and a body
consisting of a main bubble plus probably a smaller one, for a total angular
length of about $80''$.
Another record of this object is its small bow-shock size
($0.06''\pm0.02''$), marginally resolved with the Hubble Space Telescope
(Chatterjee \& Cordes \cite{cc02}), which corresponds to a linear size
$\Rbow\SIMEQ1.7\E{15}\dscal\U{cm}$; this unusually small size is a natural
consequence of the high pulsar speed.

Previous searches for an X-ray counterpart of the nebula provided negative
or, most confidently, marginal results (Romani et al.\ \cite{rcy97}).
Only an analysis of Chandra X-ray data by Hui \& Becker (\cite{hb07})
(also Cordes et al., unpublished manuscript) clearly revealed an extended
linear feature whose origin apeared to coincide with the pulsar position
and extended for over $2'$ (which at a putative distance of $1.86\U{kpc}$
implies a linear size $\Lfea$ of over 1~pc), while remaining collimated (with
a transverse size of about $20''$) and not having any apparent sign of bending.

Kargaltsev \& Pavlov (\cite{kp08}) suggested that this feature was physically
associated with another X-ray source of unknown nature.
However, even though a coincidence cannot be excluded, a physical connection
between B2224+65 and the X-ray feature appears natural.
It is supported by the fact that the linear feature points almost exactly
towards the pulsar and corresponds to a hard non-thermal spectrum, in addition
to some numerical coincidences outlined below.

Even if association with the pulsar appears likely, the nature of this X-ray
feature remains unclear.
The most puzzling fact is that its orientation deviates from the direction of
the pulsar motion by $\SIMEQ118^\circ$, and it is therefore located outside
the bow-shock region.
A scenario involving a ballistic jet does not appear viable to be an
explanation of this feature, which, although protruding for over 1~pc into
the ambient medium, remains a remarkably linear structure.
To excavate dynamically such a long path through the ambient medium without
experiencing any appreciable bending, a jet should be highly energetic and
extremely well collimated.
The requirement that the ambient medium ram pressure has not significantly
affected the direction of this ballistic motion on a scale more than 2,000
times larger than the bow-shock size (which is also affected by the same
ambient ram pressure), places an extremely tight limit on the collimation
angle of one such jet, which we estimate below.

We assume that a constant fraction $\xi$ of the pulsar spin-down power enters
an isotropic wind, while a fraction $\mu$ ($\xi+\mu<1$) is channelled into
a cylindrical jet of circular cross-section and transverse size $\Wfea$.
While the material in the jet travels a distance $\Lfea$, we also assume that
the ambient ram pressure deposits an extra transverse momentum, equal to a
factor of $\eps$ times the original jet momentum ($\eps$ must be very small,
otherwise the jet bending would be appreciable).
A comparison between the original momentum of the jet and the additional
transverse momentum implies that:
\begin{equation}
  \eps\mu\Edot/c=\Wfea\Lfea\sin(118^\circ)\rhoamb\Vpsr^2\,.
\end{equation}
The size of the bow shock is also determined by the balance between stellar
wind and ambient medium ram pressures, namely
\begin{equation}
  \xi\Edot/c=4\pi\Rbow^2\rhoamb\Vpsr^2\,.
\end{equation}
In the above two equations, $\mu$ and $\xi$ are the relative conversion
efficiencies.
Combining the above equations, we derive:
\begin{equation}
  \frac{\Wfea}{\Lfea}=\frac{4\pi\eps\mu}{\sin(118^\circ)\xi}
    \left(\frac{\Rbow}{\Lfea}\right)^2<3.6\E{-6}\frac{\eps\mu}{\xi}\,,
\end{equation}
where we have used $\Lfea>2,000\,\Rbow$.
Even though this result can be partially attenuated by assuming an ambient
density gradient, a low wind efficiency, and/or by invoking some special
transient in the pulsar energy release (even though no sign of this was
evident in the pulsar timing), it is theoretically difficult to account for
such a small $\Wfea/\Lfea$ ratio, especially considering that the measured
value of this ratio is about $0.1$--$0.2$.
Therefore, even though all the above estimates are approximate and some
assumptions could be refined, the observed value is many orders of magnitude
above the theoretical upper limit, and therefore the hypothesis of a ballistic
jet is difficult to pursue.

In this paper, we propose an alternative scenario to explain the linear X-ray
feature and its phenomenology, which is based on the idea that high-energy
electrons may diffuse away from the bow-shock region and interact with the
ambient medium.
The basic ideas behind this scenario and its assumptions are presented in
the next section.
Section~3 investigates the evolution of the highest energy electrons in the
bow-shock region, while the interaction of these electrons with the ambient
field is considered in Sect.~4.
Section~5 presents our conclusions and discusses the model predictions that
should be tested in a near future.


\section{Basic ideas and assumptions}

The principal feature of the scenario proposed here is that the highest
energy electrons accelerated at the pulsar wind termination shock may escape
from the bow-shock region and diffuse through the ambient medium, where they
emit synchrotron X-rays.
These electrons interact directly with the ambient magnetic field, and this
interaction will affect both their motion and emission.
The macroscopic dynamical properties of the ambient medium will not be
changed; for instance, the orientation of the X-ray feature simply reflects
the original orientation itself of the ambient magnetic field.
However, the electrons may play a role in the amplification of the ambient
magnetic field by creating a turbulent component, which may affect their
diffusion.
The diffusion coefficient perpendicular to the original orientation of the
magnetic field remains small, and the cross-field diffusion can be neglected.
In the following, we verify the internal consistency of these assumptions,
as well as their consistency with the observed phenomenology.

The pulsar moves with respect to the ambient medium and, as a consequence,
electrons are always injected in different flux tubes.
Even in the absence of cross-field diffusion, the nebular source has a
thickness that depends on the synchrotron lifetime of the X-ray emitting
electrons.
The map shown by Hui and Becker (\cite{hb07}) represents a transverse size of
at most $20''$; while an average transverse profile (taken from Cordes et al.,
in preparation; Romani, private communication) is consistent with a sharp
($<2''$) leading edge and a backward tail of a total thickness $\SIMEQ18''$.
A fit to this profile by an exponentially decreasing law provides an e-fold
scalelength $\SIMEQ19''$.
Using this value, the synchrotron timescale for the X-ray emitting electrons
can be estimated to be:
\begin{equation}
   \tlif=\frac{19''}{\sin(118^\circ)0.182''\U{yr^{-1}}}\SIMEQ120\U{yr}\,.
\end{equation}
Their lifetime, derived from synchrotron theory, is:
\begin{equation}
   \tlif=24.5\,B^{-2}\gm^{-1}\U{yr}
   \SIMEQ120\Escaldis^{-1/2}\Bscaldis^{-3/2}\U{yr}\,.
\end{equation}
Therefore, the synchrotron timescale as inferred from observations corresponds
to a magnetic field $\SIMEQ45\U{\mu G}$ (here and in the following, we use
$2\U{keV}$ as a reference energy for the observed X-ray photons).
With this magnetic field, the Lorentz factor of the X-ray emitting electrons
is:
\begin{equation}
   \gmX\simeq10^8\Escaldis^{2/3}\Bscaldis^{-2}\,.
\end{equation}


\section{Physical conditions in the bow-shock region}

The Lorentz factor of the electrons emitting X-rays in the nebular feature
is so large that their gyration radius is comparable with the bow-shock size.
In fact, if the magnetic field in the head of the bow shock is of the order of
the equipartition field
\begin{equation}
  \Bbow=\sqrt{2\Edot/c\Rbow^2}\SIMEQ170\,\xi^{1/2}\dscal^{-1}\U{\mu G}\,,
\end{equation}
the gyration radius of electrons with Lorentz factor $\gmX$ is
$\Rgyrbow=(m_ec^2/e\Bbow)\gmX$, namely:
\begin{equation}
  \frac{\Rgyrbow}{\Rbow}\SIMEQ0.6\,\xi^{-1/2}\Escaldis^{2/3}\Bscaldis^{-2}\,.
\end{equation}
The fact that $\Rgyrbow/\Rbow$ is close to unity supports the original
assumption that high-energy electrons may escape from the bow-shock region.
The spectrum of the X-ray feature is quite hard (as presented by Hui \&
Becker \cite{hb07}).
This implies that the electron energy distribution is dominated by its
highest-energy part, and that a leakage of the highest-energy electrons
would affect substantially the entire energy budget of the system.

Another interesting result is that the electrons produced at the termination
shock may reach quite high energies.
As a dimensional scaling, we evaluate the Lorentz factor of maximally
accelerated electrons.
According to Goldreich \& Julian (\cite{gj69}), the maximum potential drop
(between the pole and the last open field line) in an aligned pulsar is:
\begin{equation}
  \DlPhi=\left(a\Om/c\right)^2a\Bpsr=\sqrt{3\Edot/2c}\,,
\end{equation}
which corresponds to an acceleration of up to a Lorentz factor:
\begin{equation}
  \gmMPD=\frac{e\DlPhi}{2m_ec^2}=\frac{e}{2m_ec^2}\sqrt{\frac{3\Edot}{2c}}
  =7.2\E{7}\,.
\end{equation}
The processes that lead to the acceleration of high-energy electrons in pulsar
wind nebulae are complex and poorly understood, and involve both pulsar wind
acceleration and particle acceleration at the wind termination shock.
It appears that $\gmMPD$ can be used broadly as a reference value: in the Crab
Nebula, for instance, the maximum Lorentz factor of the injected electrons
is about 10\% of $\gmMPD$ (De Jager et al.\ \cite{jea96}); while in this
case a maximum energy of the injected electrons of the order of $\gmMPD$
appears to be required.


\section{Physical conditions and processes in the nebular feature}

We consider the magnetic field in the nebular feature, whose value ($45\U{\mu
G}$) was estimated from the transverse profile of the X-ray feature.
An underlying assumption was that the electrons do not diffuse orthogonally
in the ambient magnetic field.
The unavoidable jittering across the flux tube, of the order of the gyration
radius, can be estimated to be:
\begin{equation}
  \Rgyrfea
\SIMEQ0.13''\Escaldis\Bscaldis^{-3}\dscaldis^{-1}\,.
\end{equation}
This value is consistent with the upper limit of $2''$ to the leading edge of
the X-ray feature (Cordes et al., unpublished manuscript); given the strong
dependence on $B$, a magnetic field far lower than our estimate would however
imply a $\Rgyrfea$ in excess of the (measured) leading edge thickness.

Concerning the origin of this field, one cannot exclude in principle the
presence of pre-existing inhomogeneities in the ambient medium, as long as
they preserve the general plane-parallel magnetic-field structure.
A filament is clearly visible in the H$\al$ images of the Guitar Nebula,
about $100''$ to the south of and approximately parallel to the X-ray feature,
and it may be associated with a density enhancement (although it is not even
known whether it is at the same distance of the Guitar Nebula).

Another more natural possibility is that some turbulent amplification of
the originally plane-parallel field has occurred.
Some degree of turbulence would also help to explain the length of the
X-ray feature.
If electrons with lifetimes of about $120\U{yr}$ (as estimated above) flow
along the field lines close to the speed of light, they would travel for
about $36\U{pc}$; the observed length of the X-ray feature is, however, only
about $1\U{pc}$, and it is insufficient that the electrons become invisible
beyond this distance.
The X-ray luminosity of the nebular feature is extremely high,
$4.1\E{31}\dscal^2\U{\ergs}$ ($0.5$--$10\U{keV}$ band).
This is $3.5\E{-2}\dscal^2$ times the pulsar spin-down power, a high
efficiency when compared with those usually measured for pulsars ($10^{-3}$
for the pulsar+nebula X-ray emission; Becker \& Tr\"umper \cite{bt97}),
and therefore it is more natural to expect that the high-energy electrons
lose energy almost completely in the X-ray feature.

To release most of their energy in this distance, the electrons cannot flow
away freely along the flux tube.
A natural alternative is that they scatter back and forth along the flux
tube: this may be possible if, in addition to the plane-parallel large-scale
magnetic field, there is a turbulent field component.
The fact that the linear brightness of the nebular feature was observed to
decrease ``linearly'' (Hui \& Becker \cite{hb07}) may in fact be reconciled
with the case of diffusion.
In the case of steady injection and diffusion, the expected profile resembles
a decreasing exponential, and the observed profile is consistent with an
exponential decrease of e-fold scalelength of about $140''$.
In the following, we use this value for $\Lfea\sin\tau$ where, for generality,
we have also introduced the angle $\tau$ between the direction of the ambient
magnetic field and the line of sight.
Anyway, we do not assume any special field orientation in our direction,
so that $\sin\tau\sim1$.

The scattering mean free path along the field lines is:
\begin{equation}
  \lmbpar\simeq\Lfea^2/\Llif\simeq5''\sin\tau^{-2}\dscal\,.
\end{equation}
According to the diffusion theory, $\lmbpar=\eta\Rgyr$ with $\eta=(\dl
B/B)^{-2}\rs{res}$, where ``res'' means fluctuations $\delta B$ of a
wavelength that is resonant with $\Rgyr$.
Therefore, one can estimate:
\begin{eqnarray}
   \eta&\SIMEQ&37\sin\tau^{-2}\Escaldis^{-1}\Bscaldis^3\dscaldis^2\,,	\\
   \frac{\dl B}{B}&
       \SIMEQ&0.16\sin\tau\Escaldis^{1/2}\Bscaldis^{-3/2}\dscaldis^{-1}\,.
\end{eqnarray}
As required for self-consistency, $\eta$ is $>1$ and  $\dl B/B<1$.
The small value for $\dl B/B$ refers only to resonant fluctuations and
may therefore still be consistent with a global magnetic amplification
of a few times.

The derived value for $\dl B/B$ is far higher than typical values
for random fluctuations of the Galactic magnetic field with wavelengths of about
$10^{-3}\U{pc}$ (i.e.\ the gyration radius of the X-ray emitting electrons),
which is $\dl B/B\sim10^{-5}$--$10^{-4}$. We used estimates by De Marco
et al.\ (\cite{dbs07}) in the parsec range, and assumed a Kolmogorov spectrum
of fluctuations.
A possibility is that the flow of relativistic electrons itself generates,
by some streaming instability, the turbulent component of the magnetic field
responsible for the field amplification as well as the confinement of the
electrons themselves.
Without a detailed analysis of the instabilities, which is beyond
the scope of the present work, one can verify whether this scenario is
energetically consistent.
The power that feeds magnetic fields can be estimated to be:
\begin{eqnarray}
  &&(B^2/8\pi)\Rgyrfea\Lfea\sin\tau^{-1}\sin(118^\circ)\Vpsrt	\nonumber\\
  &&\quad=1.7\E{32}\sin\tau^{-1}\dscaldis^2\Escaldis^{1/3}\U{\ergs}\,,
\end{eqnarray}
which is close in value to the X-ray luminosity of the feature,
$4.1\E{31}\dscal^2\U{\ergs}$, as if an equipartition between electrons and
magnetic field has almost been established, on timescales shorter than the
synchrotron lifetime.


\section{Conclusions}

We have shown that a self-consistent scenario may exist that explains
the basic attributes of the phenomenology of the X-ray feature detected
close to the Guitar Nebula.

Some ``numerical coincidences'' between measured quantities can be readily
explained by this scenario. Its assumptions could therefore help us to
understand other peculiar features in the Guitar Nebula, and other less
extreme pulsar bow-shock nebulae, for example the possibility of the highest
energy electrons escaping from the bow-shock region.

If the magnetic field within the bow shock is close to equipartition, the
fact that the bow-shock size is similar to the gyration radius of electrons
with Lorentz factors close to $\gmMPD$ is not just a ``numerical coincidence''
valid only in the case of the Guitar Nebula.
The similarity in the two scalelengths is found for any
pulsar bow-shock nebula, independently of the pulsar spin down-power and
velocity. By substituting Eqs.\ 7 and 10 into the definition of $\Rgyr$, we
find that
\begin{equation}
  \Rgyr=\frac{m_ec^2}{e\Bbow}\gmMPD=\frac{1}{2\Bbow}\sqrt{\frac{3\Edot}{2c}}=
    \frac{\sqrt{3}}{4}\Rbow\,,
\end{equation}
The level of proximity of both the bow-shock magnetic field to its
equipartition value, and the maximum electron Lorentz factor to the value
$\gmMPD$ could produce the observed differences in the pulsar bow-shock
nebulae.
The Guitar Nebula is, however, an ideal object in which to observe these
effects, because in other bow-shock nebulae a similar X-ray feature (if scaled
to the bow-shock stand-off distance) would be too diffuse to be detectable.

From the theoretical side, a more detailed investigation is required
to understand which instabilities could allow an efficient, turbulent
amplification of the magnetic field, as required to explain the X-ray feature.
It would also be important to determine the highest energies of the electrons
injected from the wind termination shock, a long-standing problem that has
never been resolved.

From the observational side, there are some predictions of the present model
that need to be verified.
First of all, the linear feature must travel at the same speed as the pulsar:
this prediction could be tested in the next few years because the pulsar,
with its high proper motion, covers about $1''$ in only 5~years.
If no motion was be detected, a completely different scenario would have to be
envisaged.
Another prediction is the spectral change across the feature, with harder
spectra on its front side and softer spectra on its back side (due to the
synchrotron-driven evolution of the electrons, injected at different times).

Inverse Compton scattering of electrons with Lorentz factors $\sim10^8$
can upscatter CMB photons to the TeV range.
Therefore, in principle, this object could also be a TeV source, but
it would be far too faint to be detected by present day Cherenkov
telescopes.
At $B=45\U{\mu G}$, the magnetic field energy density is about a factor of
200 higher than the energy density of the CMB radiation.
Therefore, the total inverse Compton luminosity of this feature should be 200
times lower than the synchrotron luminosity, namely $\simeq2\E{29}\U{\ergs}$
corresponding to a flux $\simeq3\E{-16}\U{cm^{-2}s^{-1}}$, about three orders
of magnitude too faint to be detected by MAGIC ($5\sg$ detection in 50~hr).
Of course, detecting a TeV source at the position of the feature (its size
would allow it to be marginally resolved from the pulsar itself) would require
a revision of the above model.

Last but not least, the present model does not explain why we do not see
another X-ray feature on the opposite side of the pulsar.
In principle, electrons could flow equally well on both sides of a magnetic
flux tube.
Also, no Doppler boosting effect can be invoked because, due to scattering,
the bulk motion of the electrons in the feature is non-relativistic.

We propose that this asymmetry reflects an asymmetry in the pulsar wind
itself. Even for an axisymmetric pulsar wind, it is sufficient that the
symmetry axis of the pulsar wind is not parallel to the pulsar velocity for
such an asymmetry to be produced.
A detailed numerical modelling is required to compute quantitatively, for
different geometries, the resulting level of asymmetry in the X-ray source.


\begin{acknowledgements}
We are especially grateful to Roger Romani, for many comments and suggestions.
We also thank Elena Amato and Werner Becker for interesting discussions, as
well as Jim Cordes and Roger Romani for having disclosed to us an unpublished
manuscript containing quite interesting and inspiring observational results.
This work has been supported by contract ASI-INAF I/088/06/0, grant TH-037
(Bandiera) and by PRIN MIUR 2006, grant 02-33 (Pacini).

\end{acknowledgements}

\end{document}